\documentclass[reprint,
amsmath,amssymb, superscriptaddress
]{revtex4-1}

\usepackage[utf8]{inputenc}
\usepackage[english]{babel}
\usepackage{float}
\usepackage{xcolor}
\usepackage{graphicx}
\usepackage{dcolumn}
\usepackage{bm}
\usepackage[hidelinks]{hyperref}

\usepackage{pdfpages}

\makeatletter
\AtBeginDocument{\let\LS@rot\@undefined}
\makeatother

\hypersetup{
    colorlinks=true,
    linkcolor=blue,
    citecolor=blue,
    filecolor=blue,
    urlcolor=blue,
}

\begin{document}

\title{Optical probing of magnons and phonons  in $\rm Ni_{80}Fe_{20}$ nanodot arrays}
\author{A. Adhikari}
 \affiliation{ Department of Condensed Matter and Materials Physics, S. N. Bose National Centre for Basic Sciences,\\ Block JD, 
Sector III, Salt Lake, Kolkata 700106, India
}
\author{P. Graczyk$^{*,}$}%

\affiliation{ Institute of Molecular Physics, Polish Academy of Sciences, \\M. Smoluchowskiego 17, 60-179, Poznan, Poland
}%

\author{A. K. Chaurasiya}
\affiliation{ Department of Condensed Matter and Materials Physics, S. N. Bose National Centre for Basic Sciences,\\ Block JD, 
Sector III, Salt Lake, Kolkata 700106, India
}
\affiliation{
Department of Physics, University of Gothenburg\\
Box 100, 405 30 Gothenburg, Sweden
}

 \author{S. Mondal}
 \affiliation{ Department of Condensed Matter and Materials Physics, S. N. Bose National Centre for Basic Sciences,\\ Block JD, 
Sector III, Salt Lake, Kolkata 700106, India
}

\author{J. W. Kłos}%
\affiliation{ ISQI, Faculty of Physics and Astronomy, Adam Mickiewicz University Poznan,\\  Uniwersytetu Poznańskiego 2, 61-614, Poznań, Poland
}%
\author{A. Barman${^{*,}}$}%
 \affiliation{ Department of Condensed Matter and Materials Physics, S. N. Bose National Centre for Basic Sciences,\\ Block JD,
Sector III, Salt Lake, Kolkata 700106, India
}
 \affiliation{Department of Physics, School of Natural Sciences,\\ Shiv Nadar Institution of Eminence (Delhi NCR), Dadri UP 201314, India}
 \email{graczyk@ifmpan.poznan.pl, abarman@bose.res.in, anjan.barman@snu.edu.in}

\begin{abstract}{Control of collective spin excitations by static or dynamic strain is an emerging phenomenon that requires 
in-depth understanding 
for design of future spin-wave-regulated devices. Here, we explore mutually interacting spin waves and acoustic wave modes in addition to few non-interactive modes through all optical excitation in 
ordered arrays of $\rm Ni_{80}Fe_{20}$ nanomagnets. The acoustic wave originated from elastic deformation resonantly couple to the spin wave via magnetoelastic effect at their overlapping frequency. We demonstrate that the choice of the lattice type in which the magnetic nanodots are arranged is crucial for the observation of the magnetoelastic interaction. Therefore, the study shows that the simultaneous existence of elastic wave and spin wave offer ingeneously advantageous features to pave the way of energy-efficient magnetoacoustic devices. 
}\end{abstract}


\maketitle

\section*{\label{sec:intro}Introduction}

Spin waves (SWs) show immense potential for next-generation computing with capability of data storage, processing and transmission. Therefore, exploring new routes for energy-efficient manipulation and control of SW dynamics 
is paramount for fundamental understanding and future applications. Strain is one such pathway to control the spin degrees of freedom by the mechanical degrees of freedom\cite{2_Thevenard,3_Kovalenko,5_Dean,6_Chang,8_Casals,10_Yang,11_Sivarajah,12_De-Lin,7_Weiler}. 
Henceforth, magnetoelastic (ME) dynamics has been extensively explored in continuous magnetic media\cite{12_De-Lin,13_Weiler,14_Dreher,15_Bombeck,16_Graczyk,17_Babu} like ultrathin films, multilayers, heterostructures of highly magnetostrictive materials in addition to  fewer works on single or an array of nanomagnets\cite{18_Comin,19_Giannetti,20_Mondal, de2021resonant,21_berk,23_chiroli}.
Nanomagnet array can host topological spin states\cite{28_yu}, spin-ice states\cite{29_Chaurasiya}, complex spin textures\cite{30_Barman,31_Mondal}, and can open both magnonic\cite{32_Adhikari} and phononic\cite{33_pan,34_Zhang} band gap which can be easily tuned by various external parameters. The strain can essentially modify the magnetic landscape via lattice deformation inducing modulation in these properties of a nanomagnet. Thereby, manipulation of magnon by surface acoustic wave (SAW) in nanomagnet array of different geometrical parameters is of great importance because of the dependence of SAW frequency upon those parameters. Moreover, the impact of SAW on magnetization is assessed by the coefficient of magnetostriction of the magnetic material. 
Thin permalloy ($\rm Ni_{80}Fe_{20}$; Py hereafter) layers of nanodots exhibit a highly suppressed, but noticeable, magnetostriction coefficient. The magnetostriction depends on the film thickness, crystalline orientation as well as the crystallinity --magnetostriction has been observed both in epitaxial and polycrystalline Py films. \cite{35_Pomerantz,36_McKeehan,37_kim,38_Bozorth,Ohtani_2013, Graczyk2015}. However, this material is characterized by low damping of magnetization precession which is also important for the observation of magnetization dynamics in Py nanodots driven by SAW. 

Here, we excited and detected, in the time domain, elastic and magnetic vibrations in a two-dimensional (2D) arrays of square-shaped Py nanodots on a $\rm Si/SiO_2$ substrate using femtosecond optical pump-probe technique. We identified the elastic and magnetic modes with the help of numerical simulations. Numerical simulations reveal that the ME coupling between SAWs and SWs is possible for the square lattice of Py dots, while it is suppressed for the hexagonal lattice.

\section*{\label{sec:method} Experimental and simulation details}

\begin{figure}[!h]
\includegraphics[width=\columnwidth]{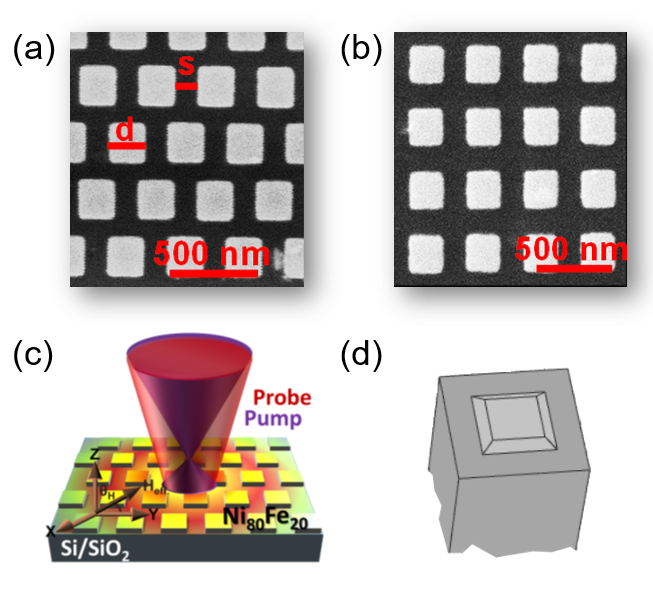}
\caption{\label{fig:structure} Scanning electron micrographs of Py nanodots forming (a) hexagonal  and  (b) square lattice. For both samples, the size of the nanodots $d=200$ nm, is fixed, and the spacing  between them $s$ is selected to set the lattice constant, $a =$ 350 nm. (c) Schematic illustration of measurement geometry used in TR-MOKE magnetometry. (d)  The geometry of the computational domain used in finite element method calculations of ME modes in  square lattice of Py dots with slanted side faces (light gray) on $\rm Si/SiO_2$ substrate.}
\end{figure}

The square-shaped nanodot arrays of Py were fabricated over 10 × 10 $\rm \mu m^2$ area on self-oxidized silicon (Si) [100] substrate using a combined process of electron-beam lithography (EBL) and electron-beam evaporation (EBE). The nanodots of width ($d$) 200 nm were arranged in hexagonal lattice (Fig.~\ref{fig:structure}(a)) or square lattice (Fig.~\ref{fig:structure}(b))  with the lattice constants 350 nm.  At first, the substrate was prepared for EBL by coating a bilayer MMA/PMMA (methyl methacrylate/polymethyl methacrylate) resist on top. During the EBL process, the beam current was 100 pA for a dose time of 1.0 $\mu$s. Next, on the resist-patterned substrate 20 nm thick Py was deposited using EBE at $1.3\times10^{-7}$~Torr base pressure. Subsequently, 5 nm thick $\rm Al_2O_3$ was deposited on top of Py to avoid possible degradation caused by impulsive laser exposure during pump-probe experiment. This was followed by a lift-off process of the resist layer. Finally, an optical quality surface was obtained for optical measurement by etching 
the residual resist using oxygen plasma.

The collective magnetization and elastic dynamics of the nanodots were measured by a home-built all-optical time-resolved magneto-optical Kerr effect (TR-MOKE) magnetometer in two colour optical pump-probe method. The second harmonic ($\lambda$ = 400 nm) of a pulsed Ti-sapphire oscillator laser was used as the pump beam to excite the dynamics, whereas the time delayed fundamental laser ($\lambda$ = 800 nm, pulse duration = 70 fs, repetition rate = 80 MHz) was used to probe the dynamics. The pump and probe beams were collinearly passed through a microscope objective of numerical aperture 0.65 for normal incidence on the sample surface. The pump beam is slightly defocused with spot diameter of $\sim$ 1000 nm while the probe beam is tightly focused to a spot diameter $\sim$ 800 nm overlapping with the pump beam. This ensures that the probed area belongs to the region uniformly excited by the Gaussian pump beam profile. Due to magneto-optical Kerr rotation, the polarization of the linearly polarized probe beam becomes elliptical. The probe beam is collected by an optical bridge detector (OBD) where the two components of the elliptically polarized light are separated to create separate electrical signals A and B. Thus, the change in Kerr rotation (differential signal: A-B) and reflectivity (sum signal: A+B) are fully isolated by the OBD and sent to two separate lock-in-amplifiers. Data are recorded simultaneously by these lock-in-amplifiers as a function of the time-delay between the pump and the probe beams. The pump beam is modulated at a frequency of 1 kHz using a mechanical chopper for phase sensitive detection used in the lock-in-amplifiers. The pump-probe measurement is carried out in polar MOKE geometry. In this experiment, a high-value bias magnetic field ($\mu_0H_{\rm ext}$=~210 mT) was first applied at a tilt angle [$\theta_H = 10^{\rm o}$ -- see Fig.~\ref{fig:structure}(c)] with the plane of the sample for in-plane saturation of the magnetization, while the tilt provides a sufficient demagnetization field in the out-of-plane direction to ensure laser-induced precessional dynamics. The magnitude of $H_{\rm ext}$ was gradually reduced in steps of almost equal interval keeping the direction fixed along the $x-$axis. At each value of $H_{\rm ext}$, the reflected probe beam was collected and detected by the OBD. During the course of the experiment, the probe fluence was fixed at 2 mJ/cm$^2$ far below the pump fluence. In our experiment, SAW is generated due to the periodic thermal expansion and contraction due to the non-uniform absorption of thermal energy produced by the high-intense optical pulses [Fig.~\ref{fig:structure}(c)]. The signature of SAW is observed in the reflectivity data as damped sinusoidal oscillation on an exponential background. A similar nature is detected in the Kerr rotation, implying the damped precessional dynamics of magnetization. The exponential background caused by the rapid increase of temperature is eliminated to extract information from the pure time domain SAW and SW dynamics.

The dynamics of SAWs and SWs can be described by a continuous model based on the elastodynamic and Landau-Lifshitz equations, respectively. The conventional magnetostatic coupling between SAWs and SWs can be expressed phenomenologically as a nonzero free energy contribution:  				
\begin{equation}
    F_{\rm me}=\frac{1}{M_{\rm s}^2}\sum_{i,j=\{x,y,z\}}\varepsilon_{ij} M_i M_j \left(b_1\delta_{ij}+b_2(1-\delta_{i,j})\right)\label{eq:me_energy}
\end{equation}
where $M_i$ are the components of magnetization vector, $\varepsilon_{ij}$ are elements of strain tensor, and $b_1$, $b_2$ denote the ME coupling constant. The coupling between the elastodynamic and Landau-Lifshitz equations is formally provided by additional contributions to dynamic effective field $\mathbf{h}_{\rm me}$ and stress tensor $\sigma_{\rm me}$ being expressed by the functional derivatives of ME free energy with respect to magnetization and strain, respectively\cite{17_Babu}. The coupled linearized Landau-Lifshitz  and elastodynamic equations take  the following form:
\begin{eqnarray}
\partial_t \mathbf{m}&=&-|\gamma|\mu_0 \left(\mathbf{M}_0\times\left(\mathbf{h}+\mathbf{h}_{\rm me}\right)-\mathbf{H}_{\rm eff}\times\mathbf{m}\right) ,\nonumber\\
\rho\partial_{t}^2u_{i}&=&\partial_k\left(\sigma_{ik}+\sigma_{{\rm me},ik}\right),\label{eq:eq_of_motion}
\end{eqnarray}
where $\mathbf{M}_0$ and $\mathbf{H}_{\rm eff}$ are equilibrium component of magnetization and static component effective field $\mathbf{H}_{\rm eff}$. The effective magnetic field $\textbf{H}_{\rm eff}(\textbf{r},t)$ consists of the external field of $H_{0}\hat{\mathbf{x}}$, the exchange field of $\textbf{H}_{\rm ex}(\textbf{r},t)= \left( 2A/\mu_{0} M^2_{\rm s} \right)\Delta\textbf{M}(\textbf{r},t)$, and the dipolar 
field of $\textbf{H}_{\rm d}(\textbf{r},t)=-\nabla \varphi (\mathbf{r},t)$. The magnetostatic potential is determined using the Gauss's law for magnetism, under the magnetostatic approximation \cite{rychly-gruszecka_shaping_2022}. The symbols $\mathbf{u}$, $\mathbf{m}$ denote the displacement vector and dynamical component of magnetization, whereas $\mathbf{h}$ stands for dynamic field and $\mathbf{\sigma}=\hat{c} \cdot \mathbf{\epsilon}$  is the stress tensor (in the absence of ME interactions) related to the strain $\mathbf{\epsilon}$ through the elastic tensor $\hat{c}$. The parameters $\rho$, $\gamma\;=196$ GHz/T, $\mu_0$ are mass density, gyromagnetic ratio and permeability of the vacuum, respectively.

%

The Eqs. (2) was implemented in COMSOL Multiphysics which used the finite element method (FEM) for numerical computations. Fig.~\ref{fig:structure}(d) presents the computational domain for square lattice of Py dots (light gray region) deposited on $\rm Si/SiO_2$ substrate (dark gray area). On the lateral boundaries of the domain, we applied the Bloch boundary conditions. For the remaining faces, we used the open boundary conditions. The height of the domain was chosen large enough to account for the exponential decay of SAW in the substrate, while at the bottom part of the substrate the perfectly matched layer is used to limit the number of bulk modes in solution. In the considered system, there is a demagnetization field caused by the presence of the lateral surfaces of the dots, which depends on their inclination. We adjusted the inclination of these walls so that the numerical model reflected the real sample and the computed SW spectrum was closest to the experimental one. The system is first relaxed at the particular external magnetic field, and then the frequency domain (linear perturbation) study is performed. In this study, the system is excited by two stimuli: a harmonic load in the form of an external pressure of amplitude $p=1$ kPa applied at the Py boundaries, and a uniform, harmonic external magnetic field of amplitude $H_{{\rm d},z}=25$ nT. For numerical calculations, we took the following values of the material parameters for Py: $\rho=8790 \text{ kg/m}^3$, $E=130$ GPa, $v=0.38$, $M_{\rm s}=760$ kA/m, $A=13$ pJ/m, $b_1=b_2=0.9$ MJ/m$^3$ \cite{Graczyk2015}, $\alpha=0.005$; SiO$_2$: $\rho=2203 \text{ kg/m}^3$, $E=73.1$ GPa, $v=0.17$; Si: $\rho=2332 \text{ kg/m}^3$, $c_{11}=166$ GPa, $c_{12}=64$ GPa, $c_{44}=80$ GPa.
The magnetoelastic coupling constants $b_1$ and $b_2$ in Py dots can attain non-negligible values due to possible nonuniform Ni–Fe composition \cite{Balakrishna_2021} and residual strain introduced during Py dot deposition. In the numerical calculations, we adopted the values used in our previous studies of a polycrystalline Py film deposited on a ferroelastic substrate, where strain-induced anisotropy gives rise to magnetostriction \cite{Graczyk2015}. 

The intensities of the modes that corresponds to the experimental intensities were calculated as  $I_p=\left| \int u_z dS \right|^2$ (for phononic signal) and $I_m=\left| \int m_z dS \right|^2$ (for the magnonic signal) over the upper boundary of the Py dot.

\section*{\label{sec:results} Results and discussion}

\begin{figure}[h]
\includegraphics[width=\columnwidth]{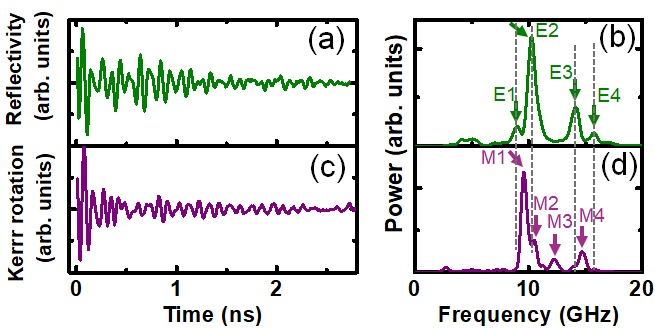}
\caption{\label{fig: TR_MOKE} Background subtracted time-resolved (a) reflectivity and (c) Kerr rotation trace indicating acoustic (SAW) and magnetization (SW) dynamics, respectively, detected for hexagonal lattice of square Py dots on SiO$_2$ substrate (see Fig.~\ref{fig:structure}(a)) at magnetic field of $\mu_0 H_{\rm ext}$~=~185~mT. (b) and (d) are FFT power spectra of the time-resolved reflectivity and Kerr rotation, respectively. The peaks indicate the frequencies of the excited SW and SAW modes, respectively.}
\end{figure}

Fig.~\ref{fig: TR_MOKE}(a) shows representative background-subtracted time-resolved reflectivity data for a hexagonal lattice at an external field of $\mu_0H_{\text{ext}}~=~185$~mT. SAW frequencies are extracted from the fast Fourier transform (FFT) of the time-domain signal, as presented in Fig.~\ref{fig: TR_MOKE}(b). Four distinct elastic modes -- E1, E2, E3, and E4 -- are observed in the 5–20 GHz range, each with varying intensity. Among these, E2 exhibits the highest amplitude, with intensities generally decreasing with increasing frequency. Notably, E1 and E2 are spectrally well-separated from E3 and E4.

\begin{figure*}[ht]
\includegraphics[width=1.0\textwidth]{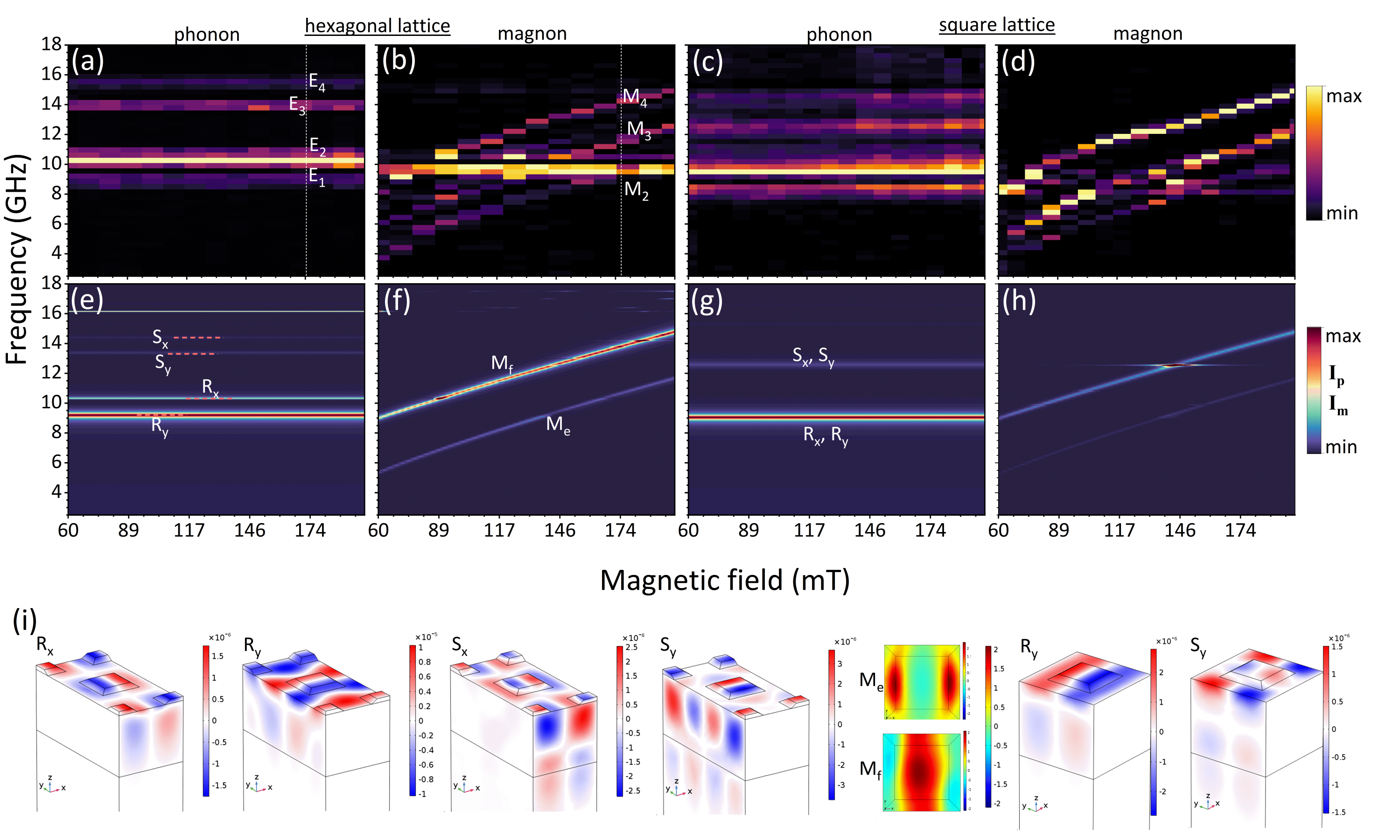}
\caption{\label{fig: spectra} The measured (a-d) and simulated (e-h) field-dependent spectra of SAW (a,c,e,g) and SW (b,d,f,h), performed for hexagonal (a,b,e,f) and square  (c,d,g,h) lattice of Py dots on ${\rm Si/SO_2}$ substrate. The simulated profiles (i) of SAW and SW allows to relate the lines in the SAW spectrum E1-E4 to Rayleigh ($\rm R_x, R_y$) and Sezawa ($\rm S_x, S_y$) SAW and the peaks in SW spectrum M3, M4 to egde mode ($\rm M_e$) and fundamental mode ($\rm M_f$) -- M2 refers to the field-independent mode. The phononic profiles shown in (i) are the displacements along the $x-$ ($\rm R_x, S_x$) or $y-$ ($\rm R_y, S_y$) direction, and the magnon profiles show the  $z-$component of magnetization. The standing SAW modes, visible at $k=0$, originate from the higher band folded into the center of the 1$^{\rm st}$ Brillouin zone. The orientation of nodal lines of SAWs on the surface indicates $x-$direction  (R$_x$, S$_x$) or $y-$direction (R$_y$, S$_y$), respectively. The profiles for SW modes show $z-$component of magnetization.} 
\end{figure*}

The emergence of multiple SAW modes arises from the periodic patterning of the SiO$_2$ substrate with Py nanodots. 
It is worth noting that periodicity causes the dispersion relation to fold within the first Brillouin zone. Using the TR-MOKE technique, one can observe only the excitations with wave vectors close to zero. This allows for the observation of standing SAWs at $k=0$ in the 1$^{\rm st}$ Brillouin zone,  originating from higher dispersion branches  with sufficiently high frequencies to interact with SWs bound in magnetic dots. 
Consequently, the lattice type and geometrical parameters of the pattern, such as the inter-dot spacing, can be used to adjust the mode frequencies, as discussed in the Supplementary Information -- Supplementary Figures~1,~2. It is worth noting that, in a continuous Py layer at $k=0$, coupling between SAW and SW modes is not possible because the SAW frequency tends to zero whereas the SW frequency remains finite. Accordingly, Supplementary Figure~3 shows that, in continuous Py, the SW mode is unaffected by the SAW.

Time-resolved Kerr rotation, shown in Fig.~\ref{fig: TR_MOKE}(c), captures the collective magnetization precession for the external field $H_{\rm ext}$ applied in $x-$direction. The corresponding FFT spectra in Fig.~\ref{fig: TR_MOKE}(d) reveal SW modes. 
At $\mu_0 H_{\text{ext}} = 185$ mT, the frequency
of magnetic modes M2 coincides with the frequency of the acoustic modes E2, suggesting a potential for ME coupling. The frequency of M1 slightly differs from its acoustic counterpart E1.

The nature of the modes becomes more prominent in the field-dependent experimental spectra shown in Fig.~\ref{fig: spectra}(a-d). SAW mode frequencies [Fig.~\ref{fig: spectra}(a,c)] remain constant across the entire $H_{\text{ext}}$ range, confirming their field-independence and the invariance of elastic properties in the nanostructure. In contrast, the magnetic spectra [Fig.~\ref{fig: spectra}(b,d)] exhibit both field-independent (M1, M2) and field-dependent (M3, M4) modes.

The field-dependent modes likely correspond to SW eigenmodes of the Py dot array (see also the change in the SW period with applied field in Supplementary Figure 4), whereas the field-independent modes originate either from forced magnetization dynamics driven by dynamic strain, or are just the reflectivity signal from modes E1, E2 that broke through to the Kerr signal. 
 For an appropriate value of the external field, $H_{\text{ext}}$, it is possible to observe an overlap between field-dependent and field-independent modes in two frequency ranges: approximately 8.5–11~GHz and 13.5–16~GHz (for hexagonal lattice -- Fig.~\ref{fig: spectra}(a)) or 7.5-10.5~GHz and 12-15~GHz (for square lattice -- Fig.~\ref{fig: spectra}(c).

M3 and M4 show a weak nonlinear field dependence on the applied field, especially at low fields, suggesting dipolar interactions. This is further evidenced by the frequency shift observed between samples with different interdot spacings $s$ (see Supplementary Figures 1, 2), proving the collective nature of these modes.

To probe the influence of lattice geometry, we examined in more details the differences between the SW spectra for a hexagonal lattice [Fig.~\ref{fig: spectra}(a,b)] and for a square array [Fig.~\ref{fig: spectra}(c,d)]  with identical lattice constant and sizes of Py nanodots. The SW spectra [field-dependent modes in Fig.~\ref{fig: spectra}(b,d)] turn out to be quite similar. This is understood as a result of weak inter-dot interaction fields in both lattices. However, SAWs appear to be more sensitive to this structural variation, with E3 and E4 frequencies significantly reduced in the square lattice. 
This is understandable because the SAWs are  delocalized and scatter on the Py dots. Therefore, the arrangement of the dots affects their frequencies. It is worth noting that the ME interaction depends on profiles of both SAWs and  SWs. Therefore, it is sufficient for only one type of excitation (in our case, SAWs) to be sensitive to the ordering of the dots for the ME interaction to be dependent on the lattice type in which the dots are arranged. The type of the lattices governs the folding of the dispersion relation and determines the spectrum of higher order standing SAW at $k=0$. The impact of the lattice is less significant for SWs localized in the Py dots-- the very flat SW dispersion is folded in a narrow frequency range.

We performed FEM simulations [Fig.~\ref{fig: spectra}(e–h)] to calculate the spatial profiles of SAWs and SWs and to study the underlying ME interactions.
In the hexagonal lattice, E1 and E2 (E3 and E4) are identified as Rayleigh R$_y$, R$_x$ (Sezawa S$_y$, S$_x$) higher order standing modes with the nodal lines in $y-$ and $x$-direction, respectively. This assignment is based on their spatial profiles [Fig.~\ref{fig: spectra}(i)]. In the square lattice [Fig.~\ref{fig: spectra}(g)], the  Rayleigh (R$_x$, R$_y$) and Sezawa (S$_x$, S$_y$) modes become degenerated and shift to lower frequencies.

On the other hand, two SW modes can be identified for both lattices: the edge mode M$\rm_e$ and the fundamental mode M$\rm_f$. While both exhibit uniform phase, M$_{\rm e}$ is localized near the lateral edges (perpendicular to $H{_\text{ext}}$), whereas M$\rm_f$ concentrates in the dot center [Fig.~\ref{fig: spectra}(i)].

We should be aware that not all modes are visible in measured and simulated TR-MOKE spectra in Fig.~\ref{fig: spectra} due to their polarization and symmetry of spatial profile. In particular, the spectra do not reveal Love SAW, which have a different polarization than Rayleigh and Sezawa waves. Moreover, due to finite damping, the frequency resolution of the TR-MOKE technique is limited by the ability to record a sufficiently long time-domain waveform for FFT analysis; hence, the direct observation of the avoided crossing remains a task for future work. The current conclusion is primarily based on a good agreement between experiment (frequency matching) and numerical calculations. In practice, frequency-resolved techniques such as BLS can mitigate this limitation, but the measurement then typically relies on driven (rather than thermally excited) magnons and phonons.

Although the experimental resolution did not allow to precisely resolve weak SW–SAW interactions, the simulations reveal notable features in the square lattice [Fig.~\ref{fig: spectra}(h)] At frequencies matching the Sezawa modes S$_x$ and S$_y$, an enhanced amplitude of magnetization dynamics is revealed in numerical outcomes -- particularly at their crossings with the fundamental SW mode M$_\mathrm{f}$. However, due to the broad linewidth of M$_\mathrm{f}$ and weak ME coupling, no clear avoided crossing appears in the experimental data.

To identify the origin of the enhanced magnetic signal in the square lattice, we conducted a detailed analysis of the ME band structure near the $\Gamma$ point, focusing on eigenfrequencies around 12.5 GHz. At this frequency, three mode types coexist: the Sezawa modes along $\tfrac{2\pi}{a}\{1,0\}$ (corresponding to S$_x$ and S$_y$), and the Love and Rayleigh modes along $\tfrac{2\pi}{a}\{1,1\}$ (with the latter denoted as R$_\mathrm{xy}$) -- see Supplementary
Information – Supplementary Figure 5. In this reciprocal lattice orientation, both Sezawa and Love modes are not expected to couple directly to M$_\mathrm{f}$, due to the known anisotropy of Rayleigh modes\cite{14_Dreher, 17_Babu}. Instead, the coupling appears to be mediated indirectly by the R$_\mathrm{xy}$ mode, which—though likely not excited or detected directly in TR-MOKE experiment -- can magnetoelastically couple with both M$_\mathrm{f}$ and the Sezawa modes in the periodic structure.

\begin{figure}[t]
\centering
\includegraphics[width=0.75\columnwidth]{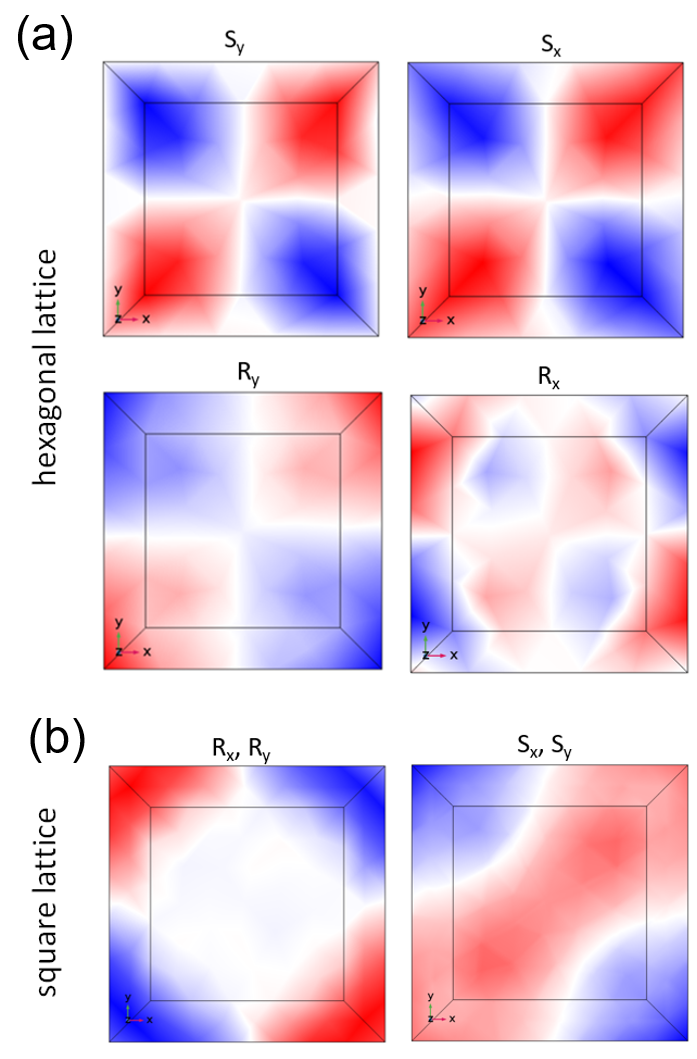} 
\caption{\label{fig:dips_S4} The spatial profile of the $y-$component of the ME field for Rayleigh ($\rm R_x, R_y$) and Sezawa ($\rm S_x, S_y$) SAW at the upper face of the Py dot, for the case of (a) hexagonal and (b) square lattice. Note that the non-zero spatial average is only observed for the Sezawa SAW in the square lattice of dots. \label{fig:Fme}}
\end{figure}

In this frequency region, the dispersion branches form a complex hybridized mode structure, complicating direct analysis of the interaction mechanisms via mode profiles. Therefore, we examined the components of the dynamic ME field at the frequencies corresponding to the Sezawa (S$_x$, S$_y$) and Rayleigh (R$_x$, R$_y$) modes. As shown in Fig.~\ref{fig:Fme}, only the $y$-component of the ME field, $h_{\mathrm{me},y}$, at the frequency of S$_x$, S$_y$ in the square lattice exhibits a non-zero spatial average over the unit cell. This component is dominated by the $\epsilon_{xy}$ element of the strain tensor. The spatial profile of $h_{\mathrm{me},y}$ aligns with the $\epsilon_{xy}$ pattern produced by the superposition of S$_x$, S$_y$, and R$_\mathrm{xy}$ modes. This result highlights the crucial role of lattice symmetry in enabling ME interactions in the system under consideration.

\section*{\label{sec:results}Conclusions}

In summary, our study revealed that femtosecond laser excitation indeed generated SAWs and SWs in the studied system. At specific field values, we observed frequency crossings between selected SAW and SW modes. These regions were investigated for signatures of dynamic ME coupling. We demonstrated that TR-MOKE measurements, combined with numerical simulations, provide a robust approach for measuring the frequencies of SAW and SW modes, identifying their polarization, and revealing dynamic ME interaction in periodic magnetic nanostructures.

Despite the common knowledge of bulk permalloy having negligible magnetostriction, which is why they are preferred in various spintronic and magnonic applications, the permalloy thin films can show sizeable magnetostriction.  Here, we studied how such undesirable, despite small, magnetostriction can affect the magnetization dynamics in magnonic crystals made of periodic array of permalloy thin film elements.

We have shown that the ME interaction for the SAW scattering on a lattice of magnetically non-interacting dots is dependent on the choice of lattice symmetry, despite the fact that the magnetization dynamics can be insensitive to the ordering of the dots. Furthermore, the interaction between SAWs and SWs in periodic magnetic nanostructures can occur for standing waves, due to the folding of the dispersion relation towards the center of the first Brillouin zone.  Higher-order standing SAWs exhibit higher frequencies, which facilitates their interaction with SWs. Lattice symmetry plays a crucial role in determining SAW frequencies and mode profiles as it dictates the folding of dispersion branches into the center of the first Brillouin zone.
Our results underscore the critical role of lattice symmetry for ME interaction in periodic magnetic nanostructures.

\section*{Author contributions}
A.B. conceived the project and, together with J.W.K., refined the initial idea. A.A.,  A.K.C. and S.M. performed the experiments. P.G. developed the theoretical model and carried out the numerical computations. A.B. supervised the work. A.A. wrote the original draft. All authors discussed the results and revised and approved the final manuscript.

\section*{Conflicts of interest}
There are no conflicts to declare.

\section*{Data availability}
The data that support the findings of this study are available from the corresponding authors upon reasonable request. The data supporting this article have been included as part of the
supplementary information (SI). \break
\noindent Supplementary Figure 1: Time-resolved MOKE spectra showing the reflectivity and Kerr rotation together with the SEM images of the three different samples S1, S2, and S3. \\
Supplementary Figure 2: The variation of SAW frequency as a function of lattice constant. \\
Supplementary Figure 3: Time-resolved MOKE showing the Kerr rotation for a continuous Py layer at different
magnetic fields.\\
Supplementary Figure 4: Magnetic field dependence of the time-resolved Kerr rotation change for sample S3.\\
Supplementary Figure 5: Magnetoelastic dispersion relation  at 143 mT for square lattice of Py dots.


\section*{Acknowledgements}

JWK acknowledges the support of the National Science Center – Poland -- grants no. UMO-2016/21/B/ST3/00452 and  no. 2020/39/O/ST5/02110 and would like to thank Grzegorz Centała for the discussion. AB gratefully acknowledges  Nano Mission, Department of Science and Technology (DST), Govt. of India (grant no. DST/NM/TUE/QM-3/2019-1C-SNB) for funding. The authors also gratefully acknowledge Professor YoshiChika Otani from The Institute for Solid State Physics, The University of Tokyo for the samples and Dr. Koustuv Dutta for assistance during the initial phase of TR-MOKE measurements.





%

\clearpage
\includepdf[pages=1]{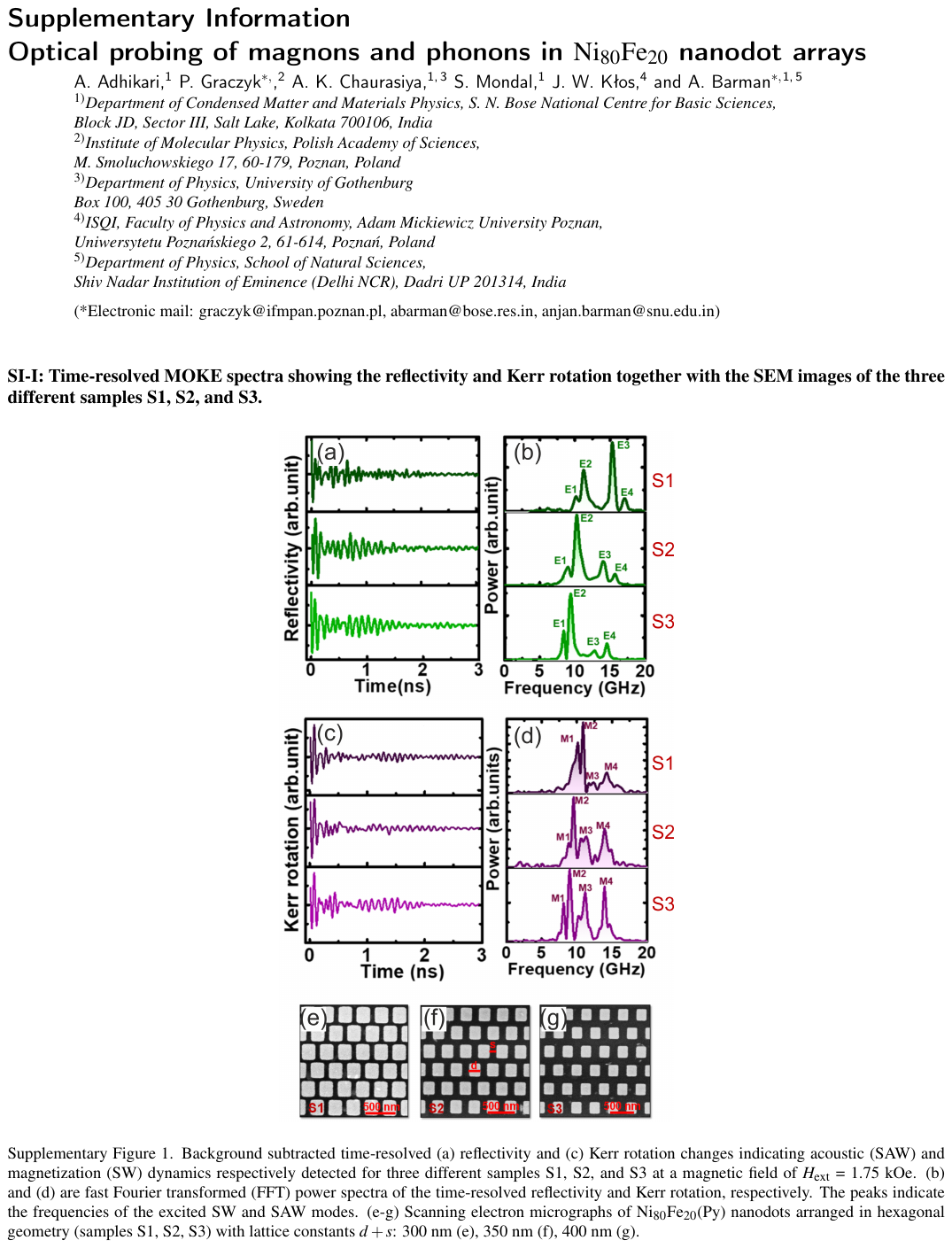}
\clearpage
\includepdf[pages=2]{SM.pdf}
\clearpage
\includepdf[pages=3]{SM.pdf}

\end{document}